\documentclass[prb,twocolumn,amsmath,amssymb,showpacs]{revtex4}

\usepackage{graphicx}
\usepackage{dcolumn}
\usepackage{bm}
\usepackage{mathrsfs}
\usepackage{appendix}
\usepackage{bbm}
\usepackage{color}

\def \vS {{\bf S}}
\def \vR {{\bf R}}
\def \ve {{\bf e}}
\def \vQ {{\bf Q}}
\def \vq {{\bf q}}
\def \vh {{\bf h}}
\def \vm {{\bf m}}
\def \vp {{\bf p}}
\def \mb {\mu_{\rm B}}

\def \pin {{\bf P}^{\rm ind} }

\def \xp {{\bf x}^{\prime }}
\def \yp {{\bf y}^{\prime }}
\def \zp {{\bf z}^{\prime }}
\def \xx {{\bf x}}
\def \yy {{\bf y}}
\def \zz {{\bf z}}

\def \vM {{\bf M}}
\def \vP {{\bf P}}
\def \BF {{\rm BiFeO$_3$} }
\def \BP {{\rm BiFeO$_3$}}
\def \TC {T_c}
\def \TN {T_{\rm N}}

\def \xq {x^{\prime}}
\def \yq {y^{\prime}}
\def \zq {z^{\prime}}

\def \Ma {{\vert \langle \delta \vert M_{\alpha } \vert 0 \rangle \vert }}

\def \cMa {\langle \delta \vert M_{\alpha } \vert 0 \rangle}
\def \cvMa { \langle \delta \vert \vM \vert 0 \rangle }

\def \Mxq {{\vert \langle \delta \vert M_{\xq } \vert 0 \rangle \vert }}
\def \Myq {{\vert \langle \delta \vert M_{\yq } \vert 0 \rangle \vert }}
\def \Mzq {{\vert \langle \delta \vert M_{\zq } \vert 0 \rangle \vert }}
\def \cMxq {\langle \delta \vert M_{\xq } \vert 0 \rangle }
\def \cMyq {\langle \delta \vert M_{\yq } \vert 0 \rangle }

\def \cPa { \langle \delta \vert P_{\alpha }^{{\rm ind}} \vert 0 \rangle }
\def \cvPa {\langle \delta \vert \vP^{{\rm ind}} \vert 0 \rangle }

\def \Pyq {{\vert \langle \delta \vert P_{\yq }^{{\rm ind}} \vert 0 \rangle \vert }}
\def \MD {{\rm MD}}
\def \ED {{\rm ED}}

\begin{document}

\title{Field dependence of the Spin State and Spectroscopic Modes of Multiferroic BiFeO$_3$}

\author{Randy S. Fishman}

\affiliation{Materials Science and Technology Division, Oak Ridge National Laboratory, Oak Ridge, Tennessee 37831, USA}

\date{\today}

\begin{abstract}

The spectroscopic modes of multiferroic BiFeO$_3$ provide detailed information about the very small anisotropy and Dzyaloshinskii-Moriya 
(DM) interactions responsible for the long-wavelength, distorted cycloid below $\TN = 640$ K.  A microscopic model 
that includes two DM interactions and easy-axis anisotropy predicts both the zero-field spectroscopic modes as well as their 
splitting and evolution in a magnetic field applied along a cubic axis.  While only six modes are optically active in zero field, 
all modes at the cycloidal wavevector are activated by a magnetic field.  The three magnetic domains of the cycloid are degenerate 
in zero field but one domain has lower energy than the other two in nonzero field.  Measurements imply that the 
higher-energy domains are depopulated above about 6 T and have a maximum critical field of 16 T, below the critical field of 19 T 
for the lowest-energy domain.  Despite the excellent agreement with the measured spectroscopic frequencies, some 
discrepancies with the measured spectroscopic intensities suggest that other weak interactions may be
missing from the model.

\end{abstract}


\maketitle

\section{Introduction}

Due to the coupling between their electric and magnetic properties, mutliferroic materials have intrigued both basic and 
applied scientists for many years.  Multiferroic materials would offer several advantages over magnetoresistive materials
in magnetic storage devices.  Most significantly, information could be written electrically and read magnetically 
without Joule heating \cite{eerenstein06}.  Hence, a material that is multiferroic at room temperature has the potential to 
radically transform the magnetic storage industry.  As the only known room-temperature multiferroic, \BF continues to 
attract intense scrutiny.

Because \BF is a ``proper" multiferroic, its ferroelectric transition temperature \cite{teague70} $\TC \approx 1100$ K is significantly higher than
its N\'eel transition temperature \cite{sosnowska82} $\TN \approx 640$ K.  Below $\TN $, a long-wavelength cycloid 
\cite{sosnowska82, rama11a, herrero10, sosnowska11} with a period of 62 nm enhances the electric polarization \cite{tokunaga10,
park11} by about 40 nC/cm$^2$.  Although much smaller than the very large polarization  \cite{lebeugle07} 
$P=100$ $\mu$C/cm$^2$ above $\TN $ but below $\TC $, the induced polarization can be used to switch between magnetic domains in an 
applied electric field \cite{lebeugle08, slee08}.  

The availability of single crystals for both elastic \cite{lebeugle08, slee08} and inelastic \cite{jeong12, matsuda12, xu12}
neutron-scattering measurements has stimulated recent progress in unravelling the microscopic interactions in \BP .  
Based on a comparison with the predicted spin-wave (SW) spectrum, inelastic neutron-scattering 
measurements \cite{jeong12, matsuda12, xu12} were used to obtain the antiferromagnetic (AF) nearest-neighbor and 
next-nearest neighbor exchanges $J_1 \approx -4.5$ meV and $J_2=-0.2$ meV, which are indicated in the 
pseudo-cubic unit cell of Fig.1(a) with lattice constant \cite{moreau71} $a \approx 3.96 \, \AA $.  When weaker interaction 
energies are suppressed by strain \cite{bai05}, non-magnetic impurities \cite{chen12}, or magnetic fields \cite{tokunaga10, park11} 
above $H_c \approx 19$ T, the exchange interactions produce a G-type AF with ferromagnetic (FM) alignment of the $S=5/2$ 
Fe$^{3+}$ spins within each hexagonal plane.  In pseudo-cubic notation, the AF wavevector is $\vQ_0 = (\pi /a)(1,1,1)$.

Below $H_c$, the much weaker anisotropy and Dzyaloshinskii-Moriya (DM) interactions produce the
distorted cycloid of bulk \BP .  For most materials with complex spin states, neutron scattering can be used to determine 
the competing interactions.   But for \BP , the cycloidal satellites at ${\bf q}=(2\pi /a)(0.5\pm \delta , 0.5, 0.5 \mp \delta )$ with $\delta \approx 0.0045$
lie extremely close to $\vQ_0$.  Because it lacks sufficient resolution in $\vq $ space, inelastic 
neutron-scattering measurements at $\vQ_0$ reveal four broad peaks below 5 meV.  Each of those peaks can be roughly 
assigned to one or more of the SW branches averaged over the first Brillouin zone \cite{matsuda12, fishman12}.

By contrast, THz spectroscopy \cite{talbayev11, nagel13} provides very precise values for the optically-active SW frequencies
at the cycloidal wavevector $\vQ $.  With polarization along $\zp =[1,1,1]$ (all unit vectors are assumed normalized to 1), 
the three magnetic domains have wavevectors $\vQ_1 = (2\pi /a )(0.5+\delta ,0.5-\delta ,0.5)$ (domain 1), 
$\vQ_2 = (2\pi /a )(0.5+\delta ,0.5,0.5-\delta )$ (domain 2), and $\vQ_3 = (2\pi /a )(0.5, 0.5+\delta ,0.5-\delta )$ (domain 3).  
The local coordinate system $\{ \xq ,\yq ,\zq \}$ for each domain is indicated in Fig.1(c).  

In zero field, the four spectroscopic modes observed below 45 cm$^{-1}$ 
were recently predicted by a model \cite{fishman13} with easy-axis anisotropy 
$K$ along $\zp $ and two DM interactions.  While the DM interaction $D$ along $\yp $ is responsible 
for the cycloidal period, the DM interaction \cite{kadomtseva04, ed05, pyatakov09, ohoyama11} $D'$ along $\zp $ 
produces the small tilt \cite{pyatakov09} $\tau $ in the plane of the cycloidal spins shown in Fig.1(b).  The tilt alternates in sign from one 
hexagonal plane to the next.  In the AF phase above $H_c$, $D'$ produces a weak FM moment \cite{tokunaga10, park11} 
perpendicular to $\zp $ due to the canting of the moments within each hexagonal plane.

This microscopic model with parameters $D$, $D'$, and $K$ also predicts the mode splitting and evolution of the spectroscopic 
modes with field.  Due to mode mixing, all of the SWs are optically active in a magnetic field.  Comparing
the predicted and observed field dependence allows us to unambiguously assign the spectroscopic modes of \BP .  
Despite the remarkable agreement between the predicted and measured mode frequencies, however, discrepancies between the 
predicted and measured spectroscopic intensities suggest that other weak interactions may be missing from the model.

We have organized this paper into five sections.  Section II discusses the spin state of \BF in a magnetic field, with results for the
wavevector, domain energies, and magnetization.  In Section III, the spectroscopic frequencies are evaluated as a function of field 
and compare those results with measurements.  The spectroscopic selection rules and intensities are discussed in Section IV.  
Section V contains a summary and conclusion.  A short description of the theory for the spectroscopic modes was recently 
presented by Nagel {\em et al.} \cite{nagel13}.

\section{Spin State}

In a magnetic field ${\bf H}=H\vm $, the spin state and SW excitations of \BF are evaluated from the Hamiltonian
\begin{eqnarray}
&&{\cal H} = -J_1\sum_{\langle i,j\rangle }\vS_i\cdot \vS_j -J_2\sum_{\langle i,j \rangle'} \vS_i\cdot \vS_j
-K\sum_i (\vS_i \cdot \zp )^2
\nonumber \\
&&-{D\, \sum}_{\vR_j=\vR_i + a(\xx -\zz )} \,\yp \cdot (\vS_i\times\vS_j) \nonumber \\
&& - {D'\, \sum}_{\vR_j=\vR_i + a\xx, a\yy, a\zz } \, (-1)^{R_{i\zq } /c} \,\zp \cdot  (\vS_i\times\vS_j)\nonumber \\
&& - 2\mb H \sum_i \vS_i \cdot \vm .
\label{Ham}
\end{eqnarray}
The nearest- and next-nearest neighbor exchange interactions $J_1=-4.5$ meV and $J_2=-0.2$ meV can be obtained from  
inelastic neutron-scattering measurements \cite{jeong12, matsuda12, xu12} between 5.5 meV and 72 meV.  
On the other hand, the small interactions $D$, $D'$, and $K$ that generate the cycloid can be obtained from spectroscopic 
measurements \cite{talbayev11, nagel13} below 5.5 meV (44.3 cm$^{-1}$).

For a given set of interaction parameters, the spin state of \BF is obtained by minimizing the energy 
$E=\langle {\cal H}\rangle $ over a set of variational parameters.  With the same spin states in hexagonal 
layers $n$ and $n+2$, the spin states in layers $n=1$ and 2 are parameterized as
\begin{eqnarray}
\label{dcx}
S_{\xq }(\vR ) &=&A^{(n)}(\vR ) \sin \mu \, \cos \tau^{(n)} \, \sin (2\pi \delta R_{\xq } /a +\gamma_1^{(n)} ) \nonumber \\
&+ & s_0 p_{\xq }^{(n)} ,
\\
\label{dcy}
S_{\yq }(\vR )& =&A^{(n)}(\vR ) \sin \mu \, \sin \tau^{(n)} \, \sin (2\pi \delta R_{\xq } /a +\gamma_2^{(n)} ) \nonumber \\
&+& s_0 p_{\yq}^{(n)} ,
\\
\label{dcz}
S_{\zq }(\vR )& =&A^{(n)}(\vR ) \cos \mu \, F^{(n)}(\vR ) + s_0 p_{\zq}^{(n)} ,
\end{eqnarray}
where
\begin{eqnarray}
&&F^{(n)}(\vR )=\sum_{l=1} C_{2l-1} \cos \bigl(2(2l-1) \pi \delta R_{\xq }/a \bigr) \nonumber \\
&&+ \sum_{l=1} C_{2l} \cos \bigl(4l \pi \delta R_{\xq }/a 
+ \Gamma^{(n)} \bigr)
\end{eqnarray}
and we take $C_1=1$.  Notice that the unit vectors $\vp^{(n)}$ and tilt angles $\tau^{(n)}$
can be different for layers 1 and 2.  Four different phases $\gamma_1^{(n)}$ and $\gamma_2^{(n)}$ 
enter $S_{\xq} (\vR)$ and $S_{\yq }(\vR )$.  In zero field, the higher odd harmonics $C_{2l+1 > 1}$ in 
$F^{(n)}(R_{\xq })$ are produced by either the anisotropy $K$ or the DM interaction $D'$.
Even harmonics $C_{2l}$ are produced by the magnetic field.  Because $C_l$ falls off rapidly with $l$,
we neglect harmonics above $l=4$.  For each layer, $\Gamma^{(n)}$ allows the even 
and odd harmonics to be out of phase.  On layer $n$ and site $\vR $, the amplitude $A^{(n)}(\vR )$ is fixed by 
the condition that $\vert \vS (\vR )\vert =S$, which is satisfied by a quadratic equation for $A^{(n)}(\vR )$.  
The lower root is used for layer 1;  the upper root is used for layer 2.

\begin{figure}
\includegraphics[width=8.5cm]{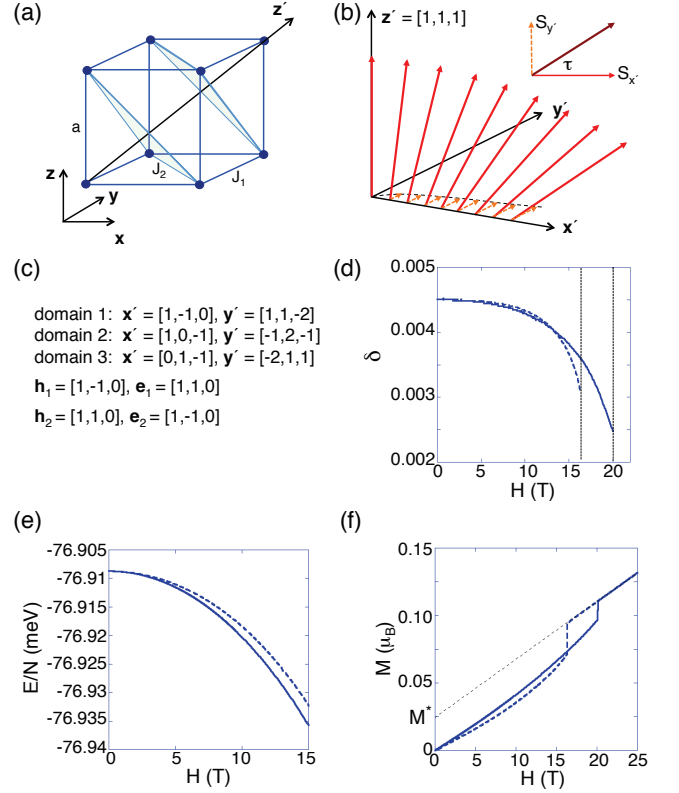}
\caption{(Color online) (a) The pseudo-cubic cell with $S=5/2$ Fe$^{3+}$ ions are at the corners.
The exchange interactions $J_1$ and $J_2$ as well as the polarization direction $\zp $ cutting through two hexagonal 
planes are indicated.  (b) For any of the three magnetic domains, 
a schematic of the spins in zero field showing their rotation about $\yp $.  Due to the DM interaction ${\bf D}'=D' \zp$, 
spins rotate by $\tau $ about $\zp $ in the $\xq \yq$ plane.  (c) The magnetic domains and THz field orientations.  
(d) The wavevector parameter $\delta $ versus field with vertical lines showing their critical fields.  
(e) The energy per site $E/N$ versus field.  (f) The magnetization $M$ along the field direction.  The thin dashed 
line shows the nonzero intercept $M^*$.  For (d), (e), and (f), the field is applied along [0,0,1], domain 1 is 
indicated by solid curves and domains 2 or 3 by dashed curves.
}
\end{figure}

Fixing $\delta  = 1/q $, where $q\gg 1 $ is an integer, $E$ is minimized over the 17 variational parameters
($\mu $, $\tau^{(n)} $, $\gamma_i^{(n)}$, $\Gamma^{(n)}$, $\vp^{(n)}$, $C_{l \le 4}$, and $s_0$) on a unit cell with 
$q$ sites along $\xp $ and two hexagonal layers.  An additional minimization loop is then performed over 
$q$ to determine the cycloidal wavevector as a function of field.  In zero field, $q=222$.
We verify that the corresponding spin state provides at least a metastable minimum of the energy $E$
by checking that the classical forces on each spin vanish.  Another check is that the 
SW frequencies are all real.

Bear in mind that the variational parameters are not free but rather are functions of the 
interaction parameters $D$, $D'$, and $K$, and the magnetic field $H$.  
In zero field, the spin state reduces to the one used in Ref.[\onlinecite{fishman13}].
A much simpler variational form for the spin state would have been possible were the field oriented along the
high-symmetry axis $\zp =[1,1,1]$ rather than along a cubic axis.

Although the number of variational parameters is far smaller than the $4q\approx 888$ degrees of freedom
for the spins in a unit cell, it may be possible to construct a more compact form for the 
spin state with fewer variational parameters.  Unlike a variational state with too few parameters,
however, a variational state with too many parameters does not incur any
penalty aside from the additional numerical expense.

With $\vm = [0,0,1]$, $\vert \vm \cdot \xp \vert$ and $\vert \vm \cdot \yp \vert $ are the same for domains 2 and 3.  
Therefore, the equilibrium and dynamical properties of domains 2 and 3 are identical.  Fig.1(d) plots $\delta $ versus field for the 
three domains.  The predicted critical field $H_c^{(2)} = 16.3$ T for domains 2 and 3 is lower than $H_c^{(1)}=20.2$ T for domain 1.
Just below $H_c^{(2)}$, the cycloid for domains 2 and 3 has a significantly longer period than the cycloid for domain 1.
The variation of $H_c^{(1)}$ with $\vm $ was predicted \cite{bras09} for a purely harmonic cycloid and 
recently reported \cite{park11} for \BP .

In zero field, all three domains have the same energy.  But in a nonzero field, domain 1 has lower energy than domains 2 and 3, 
as seen in Fig.1(e).  At 5 T, the energy difference between domains is about 0.9 $\mu $eV/site.  Based on a comparison 
between the measured and predicted spectroscopic frequencies discussed below, we conjecture that domains 2 and 3 are 
depopulated above about 6 T.

Assuming that the magnetic field is perpendicular to $\zp $,
the weak FM moment $M_0$ of the AF phase can be obtained by extrapolating the linear 
magnetization $M(H> H_c^{(2)})$ back to $H=0$.  In Ref.[\onlinecite{fishman13}], the presumed moment $M_0=0.03\mb $ 
of the AF phase was used to fix $D' = M_0 J_1/\mb S = 0.054$ meV.  For the tilted cycloid
in zero field, the spin amplitude parallel to $\yp $ is then given by $S_0=M_0/2\mb = 0.015$ and the tilt angle $\tau $ is $0.34^{\circ }$.

But neither experimental group \cite{tokunaga10, park11} applied a magnetic field perpendicular to $\zp $.  
As seen in Fig.1(f) for $\vm = [0,0,1]$ and $D'=0.054$ meV, the intercept $M^* = 0.025\mb $ is then slightly smaller than the 
measured intercept \cite{tokunaga10} $M^* =0.03\mb $.  Unlike $M_0$, $M^*(\vm )$ depends on the orientation 
$\vm $ of the magnetic field and reaches a maximum of $M_0$ only when $\vm \cdot \zp =0$ or when the field is in the $(1,1,1)$ plane.
Although a slightly larger value $D'=0.065$ meV would produce the observed $M^*(\vm )$ for $\vm = [0,0,1]$,
we retain the smaller value both because measurements of $M^*$ are rather imprecise and because
the predicted spectroscopic frequencies evaluated using $D'=0.054$ meV agree quite well with the measured frequencies.  
We shall return to this issue in the conclusion.

As also indicated in Fig.1(f), the magnetization $M(H)$ of domains 2 and 3 is lower than that of domain 1.  
A hump in the magnetic susceptibility $\chi = dM/dH $ observed \cite{park11} below 6 T may signal the 
depopulation of domains 2 and 3.

\section{Spectroscopic Frequencies}

Generally, the spin-spin correlation function $S_{\alpha \beta }(\vq ,\omega )$ may be expanded in a series of 
delta functions at each SW frequency $\omega_m (\vq )$:
\begin{equation}
\label{ssc}
S_{\alpha \beta }(\vq ,\omega )=\sum_m \delta \bigl( \omega - \omega_m (\vq ) \bigr) S_{\alpha \beta }^{(m)}(\vq ),
\end{equation}
which assumes that the SWs are not damped.  The mode frequencies $\omega_m (\vq )$
and the corresponding intensities $S_{\alpha \beta }^{(m)}$ are solved by using the $1/S$ formalism outlined in 
Ref.[\onlinecite{haraldsen09}] and in Appendix A of Ref.[\onlinecite{fishman13}].
With $\delta =1/q$, the unit cell contains $2q$ sublattices.

Some of the SW modes are optically active with non-zero magnetic dipole (MD)
matrix elements $\cvMa $, where $\vM =2\mb \sum_i \vS_i$ is the magnetization operator, $\vert 0 \rangle $ is 
the ground state with no SWs, and $\vert \delta \rangle $ is an excited state with a single SW mode at the 
cycloidal wavevector $\vQ $.  A subset of the MD modes have non-zero electric dipole (ED) matrix elements $\cvPa $,
where the induced electric polarization
\begin{equation}
\pin = {\lambda \sum}_{\vR_i, \vR_j=\vR_i + \ve_{ij} } \Bigl\{ \xp \times \bigl(\vS_i\times \vS_j\bigr) \Bigr\},
\end{equation}
of \BF is produced by the inverse DM mechanism \cite{katsura05, mostovoy06, sergienko06}.
Within each $(1,1,1)$ plane, $\ve_{ij}= \sqrt{2}a\xp $ connects spins at sites $\vR_i$ and $\vR_j$.  
In the absence of tilt, $\langle 0 \vert \vS_i \times \vS_j \vert 0 \rangle $ is parallel to $\yp $ and $\langle 0\vert  \pin \vert 0 \rangle $ 
is parallel to $\zp $.  Analytic expressions for $\cMa $ and $\cPa $ are provided in Appendix B of Ref.[\onlinecite{fishman13}].
There is no simple relationship between the SW intensities $S_{\alpha \alpha }^{(m)}(\delta )$ at the cycloidal wavevector
and the matrix elements $\cvMa $ and $\cvPa $.

For zero field with $\delta = 1/222$, we adjusted \cite{fishman13} the interaction parameters of \BF to fit the four
spectroscopic mode frequencies $\nu_0$ observed by Talbayev {\em et al.} \cite{talbayev11}.  
Fixing $D'=0.054$ meV, we obtained the parameters $D=0.107$ meV and $K = 0.0035$ meV.  We now employ 
those same parameters to describe the field dependence of the spectroscopic modes in \BP .

\begin{figure}
\includegraphics[width=9cm]{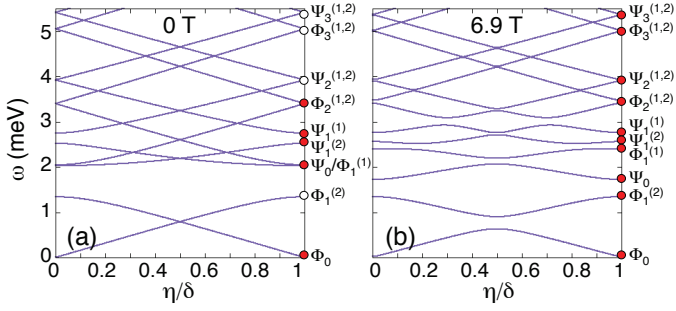}
\caption{(Color online) The mode frequencies versus ${\bf q}$  for (a) 0 T and (b) 6.9 T in domain 1.  
Optically-active modes at $\eta = \delta $ are denoted by filled circles, inactive ones by white circles.
Recall that 1 meV = 8.065 cm$^{-1}$.
}
\end{figure}

To label the spectroscopic modes at ${\bf q}={\bf Q}$ or $\eta =\delta $, we have modified the 
notation of de Sousa and Moore \cite{sousa08}, who studied the case where $D'=K=0$ so that
the cycloid is coplanar and purely harmonic.  In an extended zone scheme, they labeled the SW modes at wavevector $nQ$ 
as $\Psi_n$ and $\Phi_n$.  Corresponding to excitations within the cycloidal plane, $\Phi_n = \Phi_1\vert n\vert $ is a 
linear function of $n$.  The out-of-plane modes satisfy the relation $\Psi_n = \Phi_1 \sqrt{1+n^2}$.  
Due to the higher harmonics of the cycloid \cite{fishman12, fishman13} produced by $D'$ or $K$, 
$\Psi_n$ and $\Phi_n$ ($n > 0$) each split into two modes that we label as $\Psi_n^{(1,2)}$ and $\Phi_n^{(1,2)}$.

Any mode with a nonzero MD matrix element $\cMa $ must also have a nonzero SW intensity $S_{\beta \beta }^{(m)}(\delta )$
at the cycloidal wavevector.  When $D'=0$ and $H=0$, the cycloid is coplanar and there is a sharp distinction between in-plane and 
out-of-plane modes.  For a coplanar cycloid, the in-plane $\Phi_n$ modes may have nonzero MD matrix 
elements with component $\alpha = \yq $ and nonzero SW intensities with components $\beta = \xq$ and $\zq $.   
By contrast, out-of-plane $\Psi_n$ modes may have nonzero MD matrix elements with components $\alpha =\xq $ or $\zq $ 
and nonzero SW intensities with component $\beta =\yq $.  When $D'\ne 0$, the cycloid is tilted out of the $\xq \zq $ plane
but the distinction between the in-plane and out-of-plane modes is maintained, at least for the relatively small tilting
angles considered here:  the $\Phi_n$ modes only have SW intensities $S_{\beta \beta }^{(m)}(\delta )$ with 
$\beta = \xq$ and $\zq $ while the $\Psi_n$ modes only have SW intensity with $\beta =\yq $.  Of course, the distinction 
between in-plane and out-of-plane modes is lost in a magnetic field.

In Fig.2, the SW frequencies are plotted versus ${\bf q}=(2\pi /a) (0.5+\eta ,0.5-\eta ,0.5)$ for domain 1 and $H=0$ or 
6.9 T.  The gaps between the $\Phi_{n>0}^{(1,2)}$ and $\Psi_{n>0}^{(1,2)}$ modes at $\eta =\delta $ are 
enlarged in a magnetic field but the mode splittings fall rapidly off with increasing $n$ and cannot be seen for 
$\Phi_3^{(1,2)}$ and $\Psi_3^{(1,2)}$.  Repulsion between SW branches also occurs away from $\eta /\delta = 0$ or 1,
such as at $\eta /\delta = 1/2$.  For frequencies above a few meV, the hierarchy of modes predicted by 
de Sousa and Moore \cite{sousa08} with $\Psi_n^{(1,2)} > \Phi_n^{(1,2)}$ is restored.

As shown in Fig.2(a) for zero field,  only the six modes $\Phi_0$, $\Psi_0$, $\Phi_1^{(1)}$, $\Psi_1^{(2)}$, $\Psi_1^{(1)}$, 
and $\Phi_2^{(1)}$ are optically active at $\eta =\delta $.   At a very small but nonzero frequency, $\Phi_0$ is outside the range 
of THz measurements.  For either $D'\ne 0$ or $K\ne 0$, the anharmonicity of the cycloid splits $\Phi_1^{(2)}$ ($\nu_0=10.7$ cm$^{-1}$)
from $\Phi_1^{(1)}$ ($\nu_0=16.5$ cm$^{-1}$) and $\Psi_1^{(2)}$ ($\nu_0 = 20.4$ cm$^{-1}$) from $\Psi_1^{(1)}$  ($\nu_0 = 22.2$ cm$^{-1}$).  
Besides $\Phi_0$, only $\Psi_1^{(1)}$ has a nonzero ED matrix element in zero field.  While $\Phi_2^{(1)}$ 
($\nu_0=27.4$ cm$^{-1}$) is activated by the $3\vQ$ harmonic of the cycloid, which mixes $\Phi_2^{(1)}$ with 
$\Phi_0$, $\Psi_0$ and $\Phi_1^{(1)}$ are activated by the tilt of the cycloid out of the $\xq \zq $ plane, which mixes $\Psi_0$ with 
$\Psi_1^{(1)}$ and $\Phi_1^{(1)}$ with $\Phi_0$.   The nearly-degenerate $\Psi_0$ and $\Phi_1^{(1)}$ 
modes are responsible for the observed spectroscopic peak \cite{talbayev11, nagel13} at  $\nu_0=16.5$ cm$^{-1}$.

In nonzero field, all of the SW modes at the cycloidal wavevector $\vQ $ are optically active with nonzero MD matrix elements,
as indicated in Fig.2(b) for 6.9 T.  Notice that the near degeneracy between $\Phi_1^{(1)}$ and $\Psi_0$ is 
broken by the magnetic field.  

\begin{figure}
\includegraphics[width=10.5cm]{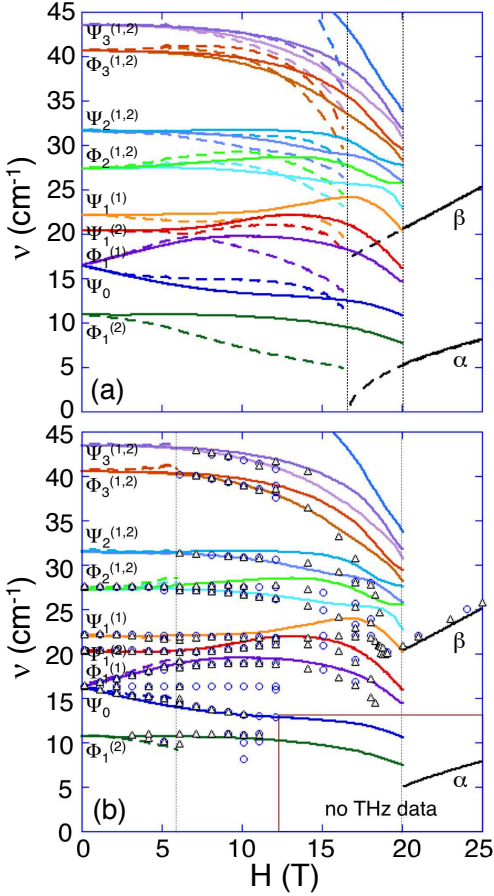}
\caption{(Color online) (a) The predicted spectroscopic frequencies for domain 1 (solid) and domains 2 and 3 (dashed).  
The critical fields are indicated by dashed vertical lines.  (b) The measured spectroscopic frequencies with 
THz field $\vh_1 $ (circles) or $\vh_2$ (triangles).  
The predicted mode frequencies from domain 1 (solid) and domains 2 and 3 (dashed) are also shown.  
We argue that contributions from domains 2 and 3 stop at 6 T, indicated by a dashed vertical line.
}
\end{figure}

With $\vm = [0,0,1]$, the predicted spectroscopic frequencies are plotted versus field in Fig.3(a) for domains 1, 2, and 3.  As
mentioned earlier, the frequencies for domains 2 and 3 are identical.  
For all three domains, $\Phi_1^{(1)}(H)$ and $\Psi_0(H)$ ($\nu_0 = 16.4$ cm$^{-1}$) are split linearly by the field below about 4 T.  
For domain 1, $\Phi_1^{(1)}(H)\approx \nu_0 + 0.9\mb H$ and $\Psi_0(H)\approx \nu_0 - 0.9 \mb H$;
for domains 2 and 3, the frequencies are slightly higher with $\Phi_1^{(1)}(H)\approx \nu_0 + 1.1\mb H$ and $\Psi_0(H)\approx \nu_0 - 0.7 \mb H$.
Some magnon softening at $\vQ $ occurs close to the critical fields $H_c^{(i)}$ for each domain.

Spectroscopic frequencies measured by Nagel {\em et al.} \cite{nagel13, miss} are plotted in Fig.3(b).  
The THz magnetic field was aligned along either $\vh_1 = [1,-1,0]$ or $\vh_2=[1,1,0]$,
with corresponding THz electric field aligned along either $\ve_1 =[1,1,0]$ or $\ve_2 = [1,-1,0]$.  These THz 
fields couple to the MD matrix elements $\langle \delta \vert \vh_i \cdot \vM \vert 0 \rangle $ and the ED
matrix elements $\langle \delta \vert \ve_i \cdot \pin  \vert 0 \rangle  $.  The observed transition 
to the AF phase occurs at about 18.9 T.  Due to instrumental limitations, no THz data is available for fields above 12 T and 
frequencies below about 12 cm$^{-1}$.  We believe that the energy difference between domains is responsible for depopulating 
domains 2 and 3 above about 6 T, indicated by a dashed vertical line.  To reflect this behavior, we have cut off the predicted mode 
frequencies of domains 2 and 3 in Fig.3(b) above 6 T.

The agreement between the measured and predicted mode frequencies in Fig.3(b) is astonishing.  
For small fields, the slopes of $\Phi_1^{(1)}(H)$ and $\Psi_0(H)$ are quite close to the predicted slopes for all three domains.
The predicted splitting of $\Phi_2^{(1,2)}(H)$ ($\nu_0 = 27.4$ cm$^{-1}$)  is clearly seen in Fig.3(b).  
Also in agreement with predictions, $\Psi_1^{(1)}(H)$ ($\nu_0 = 22.2$ cm$^{-1}$) is slightly lower in domains 2 and 3 than in domain 1.  

However, our model cannot explain the field-independent excitation at about 16.5 cm$^{-1}$ midway between $\Phi_1^{(1)}(H)$ and $\Psi_0(H)$.  
Spectroscopic modes never cross with field due to their coupling and mixing (although the coupling becomes very
weak for some higher-frequency modes).  Since it appears immune to mode repulsion, 
the 16.5 cm$^{-1}$ excitation may have some other origin, such as an optical phonon.

In contrast to the domain depopulation indicated by THz measurements, domains 2 and 3 appear to survive up to about 
16 T in electron spin resonance (ESR) measurements \cite{ruette04}.  As reported in Ref.[\onlinecite{nagel13}], 
the predicted $\Phi_1^{(2)}$ ($\nu_0 = 10.8$ cm$^{-1}$) for domains 2 and 3 agrees quite well with a mode detected by ESR measurements.

We predict that the AF phase has two low-frequency modes labeled $\alpha $ and $\beta $ in Fig.3.  
As expected, $\alpha $ and $\beta $ do not depend on the domain of the cycloid below the critical field.  
Notice that $\beta (H)$ is quite close to the Larmor frequency $2\mb H$ for an isolated spin \cite{Larmor}.  
For domains 2 and 3, $\alpha (H)$ is predicted to vanish at the critical field $H_c^{(2)}=16.3$ T.  

But ESR measurements \cite{ruette04} indicate that $\alpha (H) \approx 7.5$ cm$^{-1}$ at 16 T and that $\alpha (H)$
is projected \cite{nagel13} to vanish between 10 and 12 T.  This suggests that the true critical field $H_c^{(2)}$
for domains 2 and 3 may be as low as 10 T and that the spin state in those domains is metastable between 10 and 16 T.
Even if the critical field for domains 2 and 3 is 16 T, the depopulation of domains 2 and 3 
at 10 T would explain the optical anomalies \cite{xu09} observed at that field.  
Above $H_c^{(1)}$, $\alpha (H)\sim (H-H_c^{(2)})^{1/2}$ is sensitive to the precise location of 
$H_c^{(2)}$, which may be shifted by quantum fluctuations or other interactions not included in our model.

\section{Spectroscopic Selection Rules and Intensities}

In zero field, each optically-active mode is associated with a single MD component $\cMa $.  Besides $\Phi_0$, the
optically-active modes are:
\begin{eqnarray}
\Psi_0 &(\nu_0 = 16.4\, {\rm cm}^{-1}):  &\Mxq /\mb = 2.50 \nonumber \\
\Phi_1^{(1)}&(\nu_0 = 16.5\, {\rm cm}^{-1}): &\Myq /\mb = 1.86\nonumber \\
\Psi_1^{(2)}&(\nu_0=20.4\, {\rm cm}^{-1}): & \Mzq /\mb = 3.96\nonumber \\
\Psi_1^{(1)}&(\nu_0=22.2\, {\rm cm}^{-1}): & \Mxq /\mb = 4.59\nonumber \\
\Phi_2^{(1)}&(\nu_0 = 27.4\, {\rm cm}^{-1}): & \Myq /\mb = 1.01 \nonumber 
\end{eqnarray}
Other modes including $\Phi_1^{(2)}$ ($\nu_0=10.8$ cm$^{-1}$) and $\Phi_2^{(2)}$ ($\nu_0=27.4$ cm$^{-1}$)
are not optically active in zero field.  The only mode with a nonzero ED matrix element in zero field is 
$\Psi_1^{(1)}$ with $\Pyq /\lambda = 12.2$.

In a nonzero field, the distortion of the cycloid mixes the in-plane and out-of-plane cycloidal modes and activates all of the 
spectroscopic modes at wavevector $\vQ $.   For example, $\Phi_1^{(2)}$ ($\nu_0=10.8$ cm$^{-1}$) is not optically active and has no SW intensity 
in zero field.  But the SW intensities $S_{\alpha \alpha }(\delta )$ plotted in Fig.4(a) for domain 1
grow like $H^2$.  As shown in Fig.4(b), $\Phi_1^{(2)}$ develops significant matrix elements $\Mxq \propto H^2$ and $\Myq \propto H$.  
Despite the distortion of the cycloid in a magnetic field, $\Phi_1^{(2)}$ remains 
primarily an in-plane cycloidal mode:  $S_{\yq \yq }(\delta )$ is quite small and $\cMyq $ is the dominant MD matrix element.  But the 
significant matrix element $\cMxq $ indicates that $\Phi_1^{(2)}$ mixes with the nearby $\Psi_0$ mode.
Experimentally, $\Phi_1^{(2)}$ appears above about 3 T.

Similar conclusions hold for $\Phi_3^{(1,2)}$ ($\nu_0=40.7$ cm$^{-1}$) and $\Psi_3^{(1,2)}$ ($\nu_0=43.7$ cm$^{-1}$),
which are also activated by the field and appear above about 5 T.  The predicted splitting of both modes can be observed above 10 T.

\begin{figure}
\includegraphics[width=7cm]{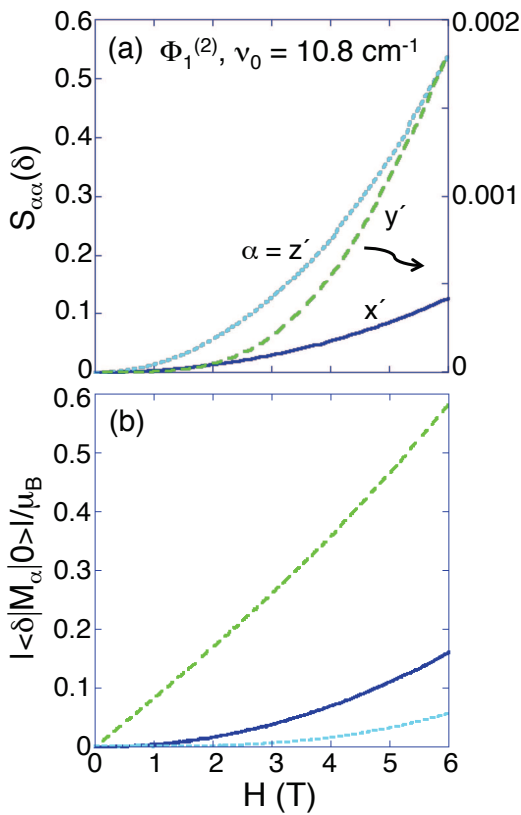}
\caption{(Color online)  The field dependence of (a) the SW intensities $S_{\alpha \alpha }(\delta )$ 
(with a different scale used for $\alpha =\yq $) and
(b) the matrix elements $\Ma /\mb $, where $\alpha = \xq $ (solid), $\yq $ (medium dash), or $\zq $ (small dash)
for $\Phi_1^{(2)}$ in domain 1.   
}
\end{figure}

Generally, the spectroscopic intensities of any mode in THz fields $\vh_i $ and $\ve_i $ ($i=1$ or 2) are given by 
\begin{equation}
\MD (\vh_i)=\Bigl\vert  \langle \delta \vert \vh_i \cdot \vM \vert 0 \rangle /\mb \Bigr\vert^2,
\end{equation} 
\begin{equation}
\ED (\ve_i)=\Bigl\vert  \langle \delta \vert \ve_i \cdot \pin \vert 0 \rangle /\lambda \Bigr\vert^2.
\end{equation} 
These expressions generalize those given in Ref.[\onlinecite{fishman13}] for zero field,
when each mode was associated with only a single matrix element $\cMa $.
The total spectroscopic intensity is a function of $\MD (\vh_i)$ and $\ED (\ve_i)$ that may also involve the
non-reciprocal cross term \cite{miyahara12} containing the product 
$\langle \delta \vert \vh_i \cdot \vM \vert 0 \rangle \, \langle 0 \vert \ve_i \cdot \pin \vert \delta \rangle $.
We expect that $\MD (\vh_i)$ dominates the spectroscopic intensity because the induced 
polarization for \BF is so small.  But measurement of non-circular magnetic dichroism \cite{miyahara12} under 
an external magnetic field along $\zp$ can, at least in principle, be used to isolate $\ED (\ve_i)$ for any mode.  

To evaluate the spectroscopic weights, we must express $\vh_i$ and $\ve_i$ in terms of the local coordinate system
$\{ x', y', z'\}$ of the cycloid in each domain:
\begin{eqnarray}
&&\vh_1=\xp , \nonumber \\
&&\vh_2= (\yp +\sqrt{2}\zp )/\sqrt{3}, 
\end{eqnarray}
in domain 1 with $\xp = [1,-1,0]$ and $\yp = [1,1,-2]$;
\begin{eqnarray}
&&\vh_1=\xp /2 -\sqrt{3}\yp /2,\nonumber \\
&&\vh_2= \xp /2 +\sqrt{3} \yp /6 +\sqrt{2/3}\zp ,
\end{eqnarray}
in domain 2 with $\xp = [1,0,-1]$ and $\yp = [-1,2,-1]$; and
\begin{eqnarray}
&&\vh_1=-\xp /2 -\sqrt{3}\yp /2, \nonumber \\
&&\vh_2= \xp /2 -\sqrt{3} \yp /6 +\sqrt{2/3}\zp ,
\end{eqnarray}
in domain 3 with $\xp = [0,1,-1]$ and $\yp = [-2,1,1]$.  For all three domains, $\ve_1=\vh_2$ and $\ve_2=\vh_1$.

\begin{figure}
\includegraphics[width=8.2cm]{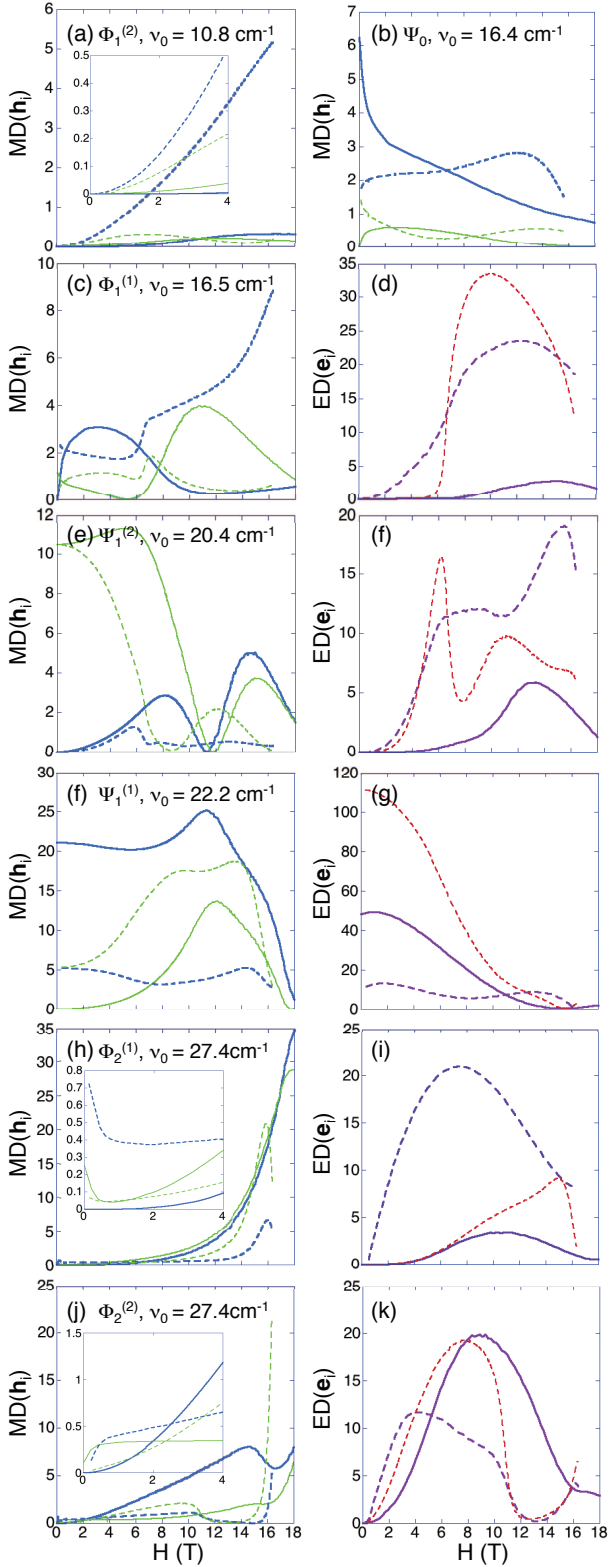}
\caption{(Color online) The spectroscopic intensities $\MD (\vh_i)$ and $\ED (\ve_i)$ versus field for the lowest 7 modes.
Domain 1 (solid) and domains 2 and 3 (dashed) are indicated along with THz fields polarizations 
$i=1$ (thick curve) and 2 (thin curve).  Side by side $\MD $ and $\ED $ plots refer to the same mode, 
indicated on the left.
}
\end{figure}

The MD and ED weights of the first seven modes above $\Phi_0$ are plotted versus field in Fig.5.  
Because they have no appreciable ED matrix elements, the ED weights of $\Phi_1^{(2)}$ and $\Psi_0$ are not shown.  
In domain 1 with $\ve_2=\xp $, $\ED (\ve_2)=0$ because $\pin $ has no component parallel to $\xp $.  
The sharp features in these figures can be attributed to the avoided crossings of the spectroscopic modes with 
field.  Experimentally, the contributions of domains 2 and 3 can be suppressed \cite{nagel13} by applying and then removing 
a high field above $H_c^{(1)}$.  

As shown in Fig.5(a) for $\Phi_1^{(2)}$ ($\nu_0 = 10.8$ cm$^{-1}$), $\MD(\vh_i )$ is much larger for domains 2 and 3 than for domain 1.  
Within domain 1, $\MD (\vh_i)$ is stronger for THz field $\vh_2$ due to the dominant matrix element $\Myq $ plotted in Fig.4(b).  
Since $\Myq $ grows linearly with field, $\MD (\vh_2)\approx \Myq^2/3$ grows quadratically with field.

In zero field, the only modes with significant ED intensity are $\Phi_0$ and $\Psi_1^{(1)}$.  By 10 T, the ED intensity
of $\Psi_1^{(1)}$ has fallen by about 66\% while the ED intensities of several other modes have become significant.
For domain 1, we predict that the ED intensity of $\Phi_2^{(2)}$ becomes comparable to 
that of $\Psi_1^{(1)}$ at about 10 T.

However, a close comparison with measurements reveals that the intensities of some activated modes
are underestimated by our model \cite{nagel13}.  For example, after averaging over domains, 
$\MD (\vh_1)$ for $\Phi_2^{(1,2)}$ is predicted to be about 25 times smaller than $\MD (\vh_1)$ for $\Psi_1^{(1)}$.
But experimentally, $\Phi_2^{(1,2)}$ has twice the intensity of $\Psi_1^{(1)}$.  For THz field orientation $\vh_1$, 
Fig.5(e) predicts that the MD intensity of $\Psi_1^{(2)}$ should vanish at $H=0$.   But experiments \cite{nagel13} indicate
that $\Psi_1^{(2)}$ survives for THz field orientation $\vh_1$ in zero field, albeit with the $\vh_1$ intensity reduced 
by about 90\% compared to the $\vh_2$ intensity.

Experimentally \cite{nagel13}, the $\vh_1$ intensities of $\Phi_2^{(1)}$ and $\Phi_2^{(2)}$ at $H=0$
are larger for the field-treated sample than for the non-field-treated sample.   
This implies that $\MD (\vh_1)$ is larger for domain 1 than for domains 2 and 3.  But the only nonzero MD
matrix element for $\Phi_2^{(1)}$ in zero field is $\cMyq $.  So as shown in 
Figs.5(h) and (j) for $\Phi_2^{(1)}$ and $\Phi_2^{(2)}$, $\MD (\vh_1) =  \Mxq^2 \rightarrow 0$ as 
$H\rightarrow 0$ in domain 1.   

\section{Conclusion}

The remarkable agreement between the predicted and measured spectroscopic mode frequencies of the cycloidal phase
leaves no doubt that a model with DM interactions along $\yp $ and $\zp$ and 
easy-axis anisotropy along $\zp $ provides the foundation for future studies of multiferroic \BP.  
However, the previous section exposed several discrepancies between the predicted and observed mode intensities which must be
addressed.  Specifically, modes that are activated by the anharmonicity and tilt of the cycloid are still too weak compared to measurements.  
Whereas our model predicts that $\Phi_2^{(1,2)}$ should not appear in zero field for domain 1 with THz field $\vh_1$, 
experiments \cite{nagel13} indicate that $\Phi_2^{(1,2)}$ are actually stronger in domain 1 than in domains 2 and 3.

As mentioned above, we have used a smaller value of $D'$ than warranted by the observed, weak FM moment $M_0$
of the AF phase.  For magnetic field along a cubic axis, $D'=0.054$ meV corresponds to 
the zero-field intercept $M^*=0.025\mb $, smaller than the intercepts $0.03\mb $ and $0.04\mb $ obtained 
by Tokunaga {\em et al.} \cite{tokunaga10} and Park {\em et al.} \cite{park11}, respectively.  
Our value $S_0=0.015$ for the cycloidal spin amplitude parallel to $\yp $ is roughly half what Ramazanoglu {\em et al.} \cite{rama11b} 
estimated from elastic neutron-scattering measurements.  Recall that the weak FM moment of the 
AF phase is predicted \cite{fishman13} to be $M_0=2\mb S_0$.  

A larger value for $D'$ requires a commensurately larger value for the anisotropy 
$K$ to preserve the same zero-field splittings of $\Phi_1^{(1,2)}$ and $\Psi_1^{(1,2)}$ produced 
by the anharmonicity of the cycloid.  For example, when $S_0=0.025$ and $D'=0.090$ meV,
the best fits to the zero-field frequencies are obtained with $K=0.0052$ meV.  
In comparison with the zero-field tilt angle $\tau  = 0.34^{\circ }$ when $S_0=0.015$, 
$\tau = 0.57^{\circ }$ when $S_0 = 0.025$.

Earlier work \cite{fishman13} found  that the matrix elements $\cMxq $ and $\cMyq $ of 
the tilt-activated modes $\Psi_0$ and $\Phi_1^{(1)}$ ($\nu_0 = 16.4$ cm$^{-1}$)
scale like $S_0$ in zero field.  So the intensities $\MD (\vh_i )$ of  $\Psi_0$ and $\Phi_1^{(1)}$ are larger by a 
factor of $25/9\approx 2.8$ for $S_0=0.025$ than for $S_0=0.015$.  But larger $D'$ and $K$ do not resolve the most serious
discrepancies between the predicted and measured intensities in zero field.  In particular, they do not
generate nonzero matrix elements $\cMxq $ for the in-plane $\Phi_2^{(1,2)}$ modes or for the out-of-plane
$\Psi_1^{(2)}$ mode at $H=0$.

Another set of weak interactions may possibly explain the enhanced spectroscopic intensities.  
There are at least two candidates for such interactions.  The small rhombohedral distortion \cite{moreau71} 
($\alpha = 89.3^\circ $) of \BF will change the next-nearest neighbor exchange $J_2$ within each hexagonal 
plane compared to the interaction between different planes.  Due to magnetoelastic coupling, easy-plane 
anisotropy perpendicular to $\yp $ may compete with the $D'$ interaction, permitting much larger values for $D'$
consistent with the observed moment $M_0$ of the AF phase.  Either set of additional interactions may modify
the MD matrix elements and change the spectroscopic intensities of the activated modes.

To conclude, the spectroscopic frequencies and intensities provide very sensitive probes of the weak
microscopic interactions that control the cycloid and induced polarization in \BP.  We are confident that future work 
based on the model presented in this paper will lay the groundwork for the eventual technological applications of this important material.

I gratefully acknowledge conversations with Nobuo Furukawa, Masaaki Matsuda, Shin Miyahara, Jan Musfeldt, 
Urmas Nagel, Satoshi Okamoto, Toomas R\~o\~om, Rogerio de Sousa, and Diyar Talbayev.  Research sponsored by the 
U.S. Department of Energy, Office of Basic Energy Sciences, Materials Sciences and Engineering Division.


\vfill

\begin{thebibliography}{}

\bibitem{eerenstein06} W. Eerenstein, N.D. Mathur, and J.F. Scott, Nat. {\bf 442}, 759 (2006).
\bibitem{teague70} J.R. Teague, R. Gerson, and W.J. James, Solid State Commun. {\bf 8}, 1073 (1970).

\bibitem{sosnowska82} I. Sosnowska, T. Peterlin-Neumaier, and E. Steichele, J. Phys. C: Solid State Phys. {\bf 15}, 4835 (1982).
\bibitem{rama11a} M. Ramazanoglu, W. Ratcliff II, Y.J. Choi, S. Lee, S.-W. Cheong, and V. Kiryukhin, Phys. Rev. B {\bf 83}, 174434 (2011).
\bibitem{herrero10} J. Herrero-Albillos, G. Catalan, J.A. Rodriguez-Velamazan, M. Viret, D. Colson, and J.F. Scott, 
J. Phys.: Condens. Matter {\bf 22}, 256001 (2010).
\bibitem{sosnowska11} I. Sosnowska and R. Przenioslo, Phys. Rev. B {\bf 84}, 144404 (2011).
\bibitem{tokunaga10}M. Tokunaga, M. Azuma, and Y. Shimakawa, J. Phys. Soc. Jpn. {\bf 79}, 064713 (2010).
\bibitem{park11} J. Park, S.-H. Lee, S. Lee, F. Gozzo, H. Kimura, Y. Noda, Y.J. Choi, V. Kiryukhin, S.-W. Cheong, Y. Jo,
E.S. Choi, L. Balicas, G.S. Jeon, and J.-G. Park, J. Phys. Soc. Jpn. {\bf 80}, 114714 (2011).
\bibitem{lebeugle07} D. Lebeugle, D. Colson, A. Forget, and M. Viret, Appl. Phys. Lett. {\bf 91}, 022907 (2007).


\bibitem{slee08} S. Lee, W.M. Ratcliff II, S.-W. Cheong, and V. Kiryukhin, Appl. Phys. Lett. {\bf 92}, 192906 (2008);
S. Lee. T. Choi, W. Ratcliff II, R. Erwin, S.-W. Cheong, and V. Kiryukhin, Phys. Rev. B {\bf 78}, 100101(R) (2008).

\bibitem{lebeugle08} D. Lebeugle, D. Colson, A. Forget, M. Viret, A.M. Bataille, and A. Gukasov, Phys. Rev. Lett. {\bf 100}, 227602 (2008).


\bibitem{jeong12} J. Jeong, E.A. Goremychkin, T. Guidi, K. Nakajima, G.S. Jeon, S.-A. Kim, S. Furukawa, Y.B. Kim,
S. Lee, V. Kiryukhin, S.-W. Cheong, and J.-G. Park, Phys. Rev. Lett. {\bf 108}, 077202 (2012).
\bibitem{matsuda12} M. Matsuda, R.S. Fishman, T. Hong, C.H. Lee, T. Ushiyama, Y. Yanagisawa, Y. Tomioka, and T. Ito, 
Phys. Rev. Lett. {\bf 109}, 067205 (2012).
\bibitem{xu12} Z. Xu, J. Wen, T. Berlijn, P.M. Gehring, C. Stock, M.B. Stone, W. Ku, G. Gu, S.M. Shapiro, R.J. Birgeneau, and G. Xu,
Phys. Rev. B {\bf 86}, 174419 (2012).

\bibitem{moreau71} J.M. Moreau, C. Michel, R. Gerson, and W.D. James, J. Phys. Chem. Sol. {\bf 32}, 1315 (1971).


\bibitem{bai05} F. Bai, J. Wang, M. Wuttig, J.F. Li, N. Wang, A.P. Pyatakov, A.K. Zvezdin, L.E. Cross, and D. Viehland, 
Appl. Phys. Lett. {\bf 86}, 032511 (2005). 
\bibitem{chen12} P. Chen, \"O. G\"unayd\i n-Sen, W.J. Ren, Z. Qin, T.V. Brinzari, S. McGill, S.-W. Cheong, and J.L. Musfeldt, 
Phys. Rev. B {\bf 86}, 014407 (2012).
\bibitem{fishman12} R.S. Fishman, N. Furukawa, J.T. Haraldsen, M. Matsuda, and S. Miyahara, Phys. Rev. B {\bf 86}, 220402(R) (2012).


\bibitem{talbayev11} D. Talbayev, S.A. Trugman, S. Lee, H.T. Yi, S.-W. Cheong, and A.J. Taylor, Phys. Rev. B {\bf 83}, 094403 (2011).
\bibitem{nagel13} U. Nagel, R.S. Fishman, T. Katuwal, H. Engelkamp, D. Talbayev, H.T. Yi, S.-W. Cheong, and T. R\~o\~om, cond.-mat.:1302.2491.
\bibitem{fishman13} R.S. Fishman, J.T. Haraldsen, N. Furukawa, and S. Miyahara, Phys. Rev. B {\bf 87}, 134416 (2013).

\bibitem{kadomtseva04} A.M. Kadomtseva, A.K. Zvezdin, Yu.F. Popv, A.P. Pyatakov, and G.P. Vorob'ev, JTEP Lett. {\bf 79}, 571 (2004).
\bibitem{ed05} C. Ederer and N.A. Spaldin, Phys. Rev. B {\bf 71}, 060401(R) (2005).
\bibitem{pyatakov09} A.P. Pyatakov and A.K. Zvezdin, Eur. Phys. J. B {\bf 71}, 419 (2009).
\bibitem{ohoyama11} K. Ohoyama, S. Lee, S. Yoshii, Y. Narumi, T. Morioka, H. Nojiri, G.S. Jeon, S.-W. Cheong, and J.-G. Park, J. Phys. Soc. Jpn. {\bf 80}, 125001 (2011).
\bibitem{bras09} G. Le Bras, D. Colson, A. Forget, N. Genand-Riondet, R. Tourbot, and P. Bonville, Phys. Rev. B {\bf 80}, 134417 (2009).
\bibitem{haraldsen09} J.T. Haraldsen and R.S. Fishman, J. Phys.:  Condens. Matter {\bf 21}, 216001 (2009).

\bibitem{katsura05}H. Katsura, N. Nagaosa, and A.V. Balatsky, Phys. Rev. Lett. {\bf 95}, 057205 (2005).
\bibitem{mostovoy06}M. Mostovoy, Phys. Rev. Lett. {\bf 96}, 067601 (2006).
\bibitem{sergienko06} I.A. Sergienko and E. Dagotto, Phys. Rev. B {\bf 73}, 094434 (2006).


\bibitem{sousa08} R. de Sousa and J.E. Moore, Phys. Rev. B {\bf 77}, 012406 (2008).

\bibitem{miss} The data plotted in Fig.3(b) does not include the observed $H=0$ modes \cite{nagel13} 
at 18.1 and 18.2 cm$^{-1}$ because they only appear in zero field.
\bibitem{ruette04} B. Ruette, S. Zvyagin, A.P. Pyatakov, A. Bush, J.F. Li, V.I. Belotelov, A.K. Zvezdin, and D. Viehland, Phys. Rev. B {\bf 69},
064114 (2004).

\bibitem{Larmor} Earlier work \cite{matsuda12, fishman12, fishman13} adopted the convention of multiplying the SW frequencies by $\sqrt{S(S+1)}$
instead of by $S$.  This scaling produces the unphysical result that the Larmor frequency for an isolated spin is $\sqrt{(S+1)/S}2 \mb H$
rather than $2\mb H$.  To rectify this, the frequency $\beta $ is scaled by $S$ rather than by $\sqrt{S(S+1)}$ in the AF phase.


\bibitem{xu09} X.S. Xu, T.V. Brinzari, S. Lee, Y.H. Chu, L.W. Martin, A. Kumar, S. McGill, R.C. Rai, R. Ramesh, V. Gopalan, S.-W. Cheong, and
J.L. Musfeldt, Phys. Rev. B {\bf 79} 134425 (2009).

\bibitem{miyahara12} S. Miyahara and N. Furukawa, J. Phys. Soc. Jpn. {\bf 80}, 073708 (2011);  {\it ibid.} {\bf 81}, 023712 (2012).

\bibitem{rama11b} M. Ramazanoglu, M. Laver, W. Ratcliff II, S.M. Watson, W.C. Chen, A. Jackson, K. Kothapalli, S. Lee, S.-W. Cheong, 
and V. Kiryukhin, Phys. Rev. Lett. {\bf 107}, 207206 (2011).


\end{thebibliography}
\end{document}